\documentclass[
reprint,
superscriptaddress,
frontmatterverbose, 
 amsmath,amssymb,
 aps,
prb,
]{revtex4-2}

\usepackage{graphicx}% Include figure files
\usepackage{dcolumn}% Align table columns on decimal point
\usepackage{bm}% bold math
\usepackage{amsmath}
\usepackage{amssymb}
\usepackage{CJK}
\usepackage{keyval}
\usepackage{color}
\usepackage{txfonts}
\usepackage{verbatim}
\begin{document}

\preprint{APS/123-QED}

\title{Interface suppressed nematicity and enhanced superconductivity of FeSe/NdFeO{$_3$} in the low doping regime}

\author{Chihao Li}
\affiliation{Laboratory of Advanced Materials, State Key Laboratory of Surface Physics, \\and Department of Physics, Fudan University, Shanghai 200438, China}

\author{Yuanhe Song}
\affiliation{Laboratory of Advanced Materials, State Key Laboratory of Surface Physics, \\and Department of Physics, Fudan University, Shanghai 200438, China}

\author{Xiaoxiao Wang}
\affiliation{Laboratory of Advanced Materials, State Key Laboratory of Surface Physics, \\and Department of Physics, Fudan University, Shanghai 200438, China}

\author{Minyinan Lei}
\affiliation{Laboratory of Advanced Materials, State Key Laboratory of Surface Physics, \\and Department of Physics, Fudan University, Shanghai 200438, China}

\author{Xiaoyang Chen}
\affiliation{Laboratory of Advanced Materials, State Key Laboratory of Surface Physics, \\and Department of Physics, Fudan University, Shanghai 200438, China}

\author{Haichao Xu}
\email{xuhaichao@fudan.edu.cn}
\affiliation{Laboratory of Advanced Materials, State Key Laboratory of Surface Physics, \\and Department of Physics, Fudan University, Shanghai 200438, China}
\affiliation{Shanghai Research Center for Quantum Sciences, Shanghai 201315, China}

\author{Rui Peng}
\email{pengrui@fudan.edu.cn}
\affiliation{Laboratory of Advanced Materials, State Key Laboratory of Surface Physics, \\and Department of Physics, Fudan University, Shanghai 200438, China}
\affiliation{Shanghai Research Center for Quantum Sciences, Shanghai 201315, China}

\author{Donglai Feng}
\email{dlfeng@ustc.edu.cn}
\affiliation{Laboratory of Advanced Materials, State Key Laboratory of Surface Physics, \\and Department of Physics, Fudan University, Shanghai 200438, China}
\affiliation{National Synchrotron Radiation Laboratory and School of Nuclear Science and Technology, University of Science and Technology of China, Hefei, 230026, China}
\affiliation{New Cornerstone Science Laboratory, University of Science and Technology of China, Hefei, 230026, China}
\affiliation{Collaborative Innovation Center of Advanced Microstructures, Nanjing 210093, China}

\date{\today}% It is always \today, today,
             %  but any date may be explicitly specified

\begin{abstract}
The discovery of interface-enhanced superconductivity in single-layer FeSe/oxides has generated intensive research interests. Beyond the family of FeSe interfaced with various TiO$_2$ terminated oxides, high pairing temperature up to 80~K has been recently observed in FeSe interfaced with FeO$_x$-terminated LaFeO$_3$. Here we successfully extend the FeSe/FeO$_x$ superconducting interface to FeSe/NdFeO$_3$, by constructing 1uc-FeSe/6uc-NdFeO$_3$/Nb:SrTiO$_3$ heterostructures. Intriguingly, well-annealed FeSe/NdFeO$_3$ exhibits a low doping level of 0.038$\sim$0.046   ~e$^-/$Fe which deviates universally magic doping level (0.10$\sim$0.12 e$^-/\rm{Fe}$) and provides a new playground for studying the FeSe/oxide interface in the low electron-doped regime. Comparing it with thick FeSe films at the comparable electron doping level induced by surface potassium dosing, FeSe/NdFeO$_3$  shows a larger superconducting gap and the absence of a nematic gap, indicating an enhancement of the superconductivity and suppression of nematicity by the FeSe/FeO$_x$ interface. These results not only expand the FeSe/FeO$_x$ superconducting family but also enrich the current understanding on the roles of the oxide interface.

\end{abstract}

%\keywords{Suggested keywords}%Use showkeys class option if keyword
                              %display desired
\maketitle

%\tableofcontents
\section{I\MakeLowercase{ntroduction}}

The discovery of high-temperature interfacial superconductivity in one unit cell (1uc) FeSe/SrTiO$_3$ \cite{wang2012interface} has ignited intensive research interests. The superconducting pairing temperature in  1uc FeSe/SrTiO$_3$ (T$_g$ $\sim$ 65K) \cite{he2013phase,tan2013interface} exhibits a substantial enhancement compared to that of bulk FeSe crystals  (superconducting critical temperature T$_c$ $\sim$ 8K) \cite{hsu2008superconductivity}. Interface engineering on the expitaxial strain, charge transfer, and interfacial electron-phonon coupling have been conducted \cite{sobota2021angle,peng2014tuning,peng2014measurement,wen2016anomalous,song2019evidence,faeth2021interfacial,peng2014measurement,peng2014tuning,zhang2017ubiquitous,zhang2017origin,lee2014interfacial,rebec2017coexistence,tang2016interface}, revealing the significant contributions of interface charge transfer and interfacial electron-phonon coupling to the enhancement of superconducting pairing \cite{faeth2021interfacial,zhang2017ubiquitous,song2019evidence,zhang2017origin,lee2014interfacial,rebec2017coexistence,tang2016interface}.

To examine the generality of the mechanism underlying the enhancement of interfacial superconductivity and to explore new interfacial superconductors, it is crucial to investigate FeSe films interfaced with other oxides. 
Extending the superconducting interfaces, a range of FeSe interfaces with TiO$_x$-terminated oxides has been explored.
These include 1uc FeSe on  BaTiO$_3$ \cite{peng2014tuning}, on anatase TiO$_2$ (001) \cite{ding2016high}, on rutile TiO$_2$ (100) \cite{rebec2017coexistence}, SrTiO$_3$ (110) \cite{zhang2017ubiquitous,zhou2016interface,zhang2016observation}, on EuTiO$_3$ \cite{liu2021high} and on LaTiO$_3$ \cite{jia2021magic}. 
Angle-resolved photoemission spectroscopy (ARPES) measurements have revealed that well-annealed 1uc FeSe on different TiO$_x$-terminated interfaces generally exhibit substantial electron doping (0.10$\sim$0.12 e$^-/\rm{Fe}$),
referred as “magic doping” \cite{li2022modulation,jia2021magic}). It puts 1~uc FeSe/titanate at the heavy electron doping regime of the phase diagram, where nematicity is absent \cite{tan2013interface,peng2014measurement,peng2014tuning,wen2016anomalous,zhang2016effects}. Moreover, replica bands reflecting the interfacial electron-phonon coupling are commonly observed in these FeSe-oxide interfaces \cite{lee2014interfacial,rebec2017coexistence,li2018electron,faeth2021interfacial,zhang2017ubiquitous}. However,  it remains challenging to construct new types of FeSe/oxide interfaces beyond the FeSe/TiO$_x$ structure in order to test the generality of this mechanism.

\begin{figure*}[bht]
\includegraphics{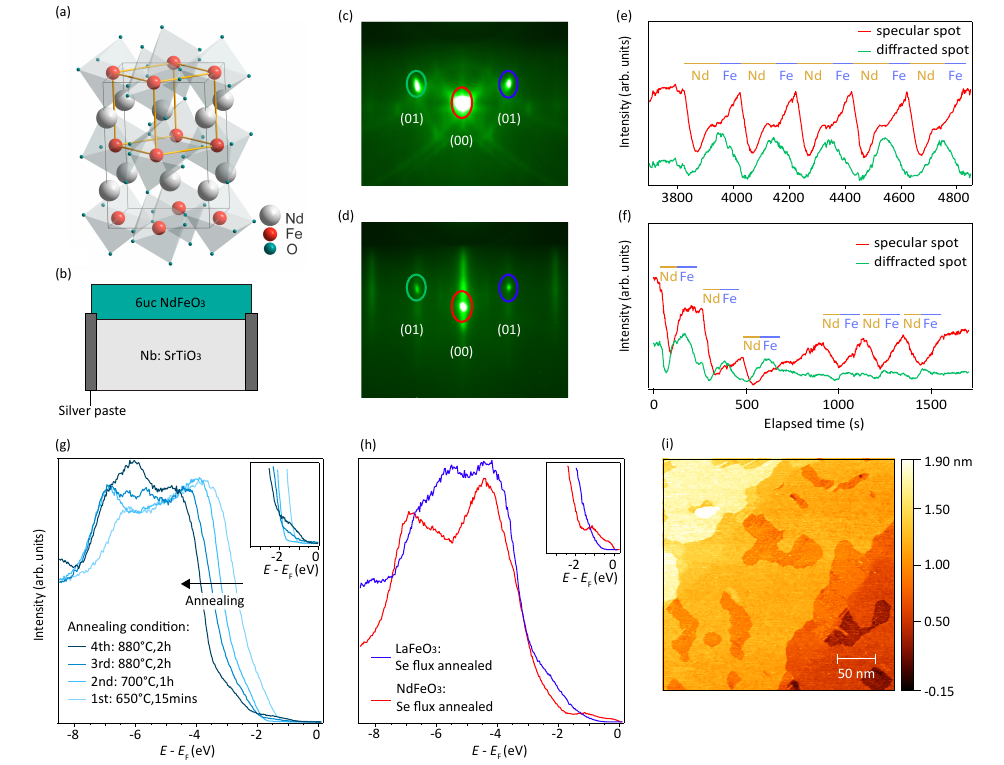}% Here is how to import EPS art
\caption{\label{Figure1}(a) Crystal structure of NdFeO$_3$. (b) Schematic illustration of the heterostructure. (c) Reflection high-energy electron diffraction (RHEED) pattern of the annealed Nb:SrTiO$_3$ substrates, showing sharp Kikuchi lines. (d) RHEED pattern of the epitaxial 6~uc NdFeO$_3$ films, showing two-dimensional RHEED streaks. (e) The intensity oscillation of the diffracted and specular RHEED spots during the growth of a calibration sample. (f) The intensity oscillation of the diffracted and specular RHEED spots during the growth of the 6~uc NdFeO$_3$. 
(g) Valence band comparison on the 6~uc NdFeO$_{3-\delta}$/Nb:SrTiO$_3$ film after each annealing experiment. The valence bands shift to higher binding energy with longer annealing times, suggesting increasing electron doping due to oxygen vacancies. (h) Valence band comparison between Se flux-annealed 6-uc NdFeO$_{3-\delta}$/Nb:SrTiO$_3$ and Se flux-annealed 6-uc LaFeO$_3$/Nb:SrTiO$_3$ \cite{song2021high}. They exhibit comparable oxygen vacancy density as indicated by the valence band position.
(i) Scanning Tunneling Microscopy (STM) image of the annealed 6uc-NdFeO$_3$/Nb:SrTiO$_3$.}
\end{figure*}

Very recently, Song et.~al reported the successful fabrication of high-quality monolayer FeSe films on the FeO$_x$-terminated LaFeO$_3$ \cite{song2021high}, in which the superconducting pairing temperature reaches 80~K, surpassing the boiling temperature of nitrogen. This work suggests that a similar pairing enhancement mechanism operates at the FeSe/LaFeO$_3$ interface as observed in the FeSe/TiO$_x$-terminated interfaces \cite{song2021high}. Yet, the FeSe/LaFeO$_3$ interface remains the only reported instance of the FeSe/FeO$_x$ superconducting interface.  Could it be possible to achieve epitaxial growth of 1uc-FeSe on other XFeO$_3$ (X=Nd, Pr, Bi, and etc.) perovskites? By expanding the FeSe/FeO$_x$ family, we can potentially establish new platforms to explore the characters of FeSe-based interfacial superconductivity. In particular, it is worth noting that the reported FeSe/LaFeO$_3$ exhibits a lower electron doping (0.087 e$^-/\rm{Fe}$) compared to the magic doping observed in FeSe/TiO$_x$-terminated interfaces. By studying FeSe on various XFeO$_3$ substrates, valuable insights can be gained into the FeSe/oxide interface under doping conditions that differ from FeSe/titanates. 

Here we have successfully fabricated a new high-quality FeSe/FeO$_x$ interface by growing a 1uc FeSe film on a 6uc epitaxially grown NdFeO$_3$ layers on Nb:SrTiO$_3$ substrates. Through in-situ ARPES measurements, we have determined that the electron doping level of the well-annealed 1uc FeSe/NdFeO$_3$/Nb:SrTiO$_3$ interface is only 0.038$\sim$0.046 e$^-/\rm{Fe}$, which is lower than the doping levels observed in the various 1uc FeSe on TiO$_x$-terminated oxides (0.10$\sim$ 0.12e$^-/\rm{Fe}$) \cite{tan2013interface,peng2014measurement,peng2014tuning,ding2016high,rebec2017coexistence,liu2021high,jia2021magic} and in FeSe/LaFeO$_3$ (0.087 e$^-/\rm{Fe}$) interfaces \cite{song2021high}, placing the superconducting interface  of FeSe/NdFeO$_3$ in the low doping regime of electron-doped FeSe phase diagram.  Compared to FeSe thick films with similar levels of electron doping  \cite{wen2016anomalous,zhang2016effects,song2016observation}, the FeSe/NdFeO$_3$ system exhibit an enhancement of superconductivity and the suppression of nematic splitting. This observation suggests that the FeSe/FeO$_x$ interface plays an additional role beyond the electron doping.
Our findings not only extend the FeSe/FeO$_x$ superconducting family, but also provide evidence for the tailored band structure achieved through the interface, shedding light on the intricate interplay between interfacial superconductivity and nematicity.

\section{MBE \MakeLowercase{growth of 1uc} F\MakeLowercase{e}S\MakeLowercase{e}/6\MakeLowercase{uc} N\MakeLowercase{d}F\MakeLowercase{e}O$_3$/N\MakeLowercase{b}:S\MakeLowercase{r}T\MakeLowercase{i}O$_3$}

NdFeO$_3$ (NFO) bulk crystals exhibit a structurally distorted perovskite structure with lattice constants of $a$=5.450~$\AA$, $b$=5.587~$\AA$, $c$=7.761~$\AA$ [Fig. \ref{Figure1}(a)]~\cite{koehler1960neutron}. The pseudo-cubic lattice constant of NFO is $a'$=3.895~$\AA$  and the lattice mismatch with SrTiO$_3$ substrate is only -0.25$\%$. Commercial  TiO$_x$ terminated 0.5$\%$wt Nb-doped SrTiO$_3$(001) substrates (NSTO) with 0.1$^\circ$ $\sim$ 0.15 $^\circ$  miscut angle (CrysTec) were used as substrates. The NSTO substrates were annealed at 500$^\circ$C in the molecular beam epitaxy (MBE) chamber (base vacuum $7 \times 10 ^{-10}$ mbar) for 1 hour, using an pressure of $2 \times 10 ^{-7}$ mbar under ozone atmosphere. The RHEED pattern of the annealed NSTO substrates shows sharp Kikuchi lines  [Fig. \ref{Figure1}(c)]. Subsequently,  NFO films were grown on the annealed substrates  by ozone-assisted molecular beam epitaxy at 650$^\circ$C with pressure of $2 \times 10 ^{-7}$ mbar under ozone atmosphere.  By alternately opening the pneumatic shutters of the Nd source and the Fe source, the oxidized Nd and Fe layers were alternately grown on the substrate in an atomic layer-by-layer mode. The relative stoichiometry
and the absolute amount of each element (La and
Fe) was first calibrated by quartz crystal microbalance (QCM) measurements and then by real-time intensity oscillations of reflection high-energy electron diffraction (RHEED) during the growth 
\cite{haeni2000rheed,peng2014measurement,peng2014tuning}. After calibration, stable oscillations of RHEED intensity can be obtained, indicating a stoichiometric Nd-Fe atomic flux ratio [Fig.\ref{Figure1} (e)]. 
Subsequently, 6~uc NdFeO$_3$ films were grown on freshly-annealed Nb:SrTiO$_3$ substrates following the calibrated stoichiometry [Fig.~\ref{Figure1}(f)]. The RHEED streaks suggest two-dimensional growth character of the grown NdFeO$_3$ films  [Fig. \ref{Figure1}(d)].  Using this atomic layer-by-layer epitaxial growth method, we control  NdFeO$_3$ to be FeO$_x$-terminated. 

Oxygen vacancies in the substrates are vital for epitaxially growing high-quality FeSe films 
\cite{wang2012interface,tan2013interface,song2019evidence,gong2019oxygen,o2022imaging,jia2021magic,ding2016high,rebec2017coexistence,zhang2017ubiquitous,zhou2016interface,zhang2016observation,liu2021high}. 
To induce sufficient oxygen vacancies, the NdFeO$_3$ films were annealed four times under vacuum, and the valence bands were examined using ARPES after each annealing step. As shown in Fig.~\ref{Figure1}(g), the valence bands (VB) of NdFeO$_{3-\delta}$/Nb:SrTiO$_3$ shift towards higher binding energy upon vacuum annealing, suggesting increasing electron doping with increasing oxygen vacancies. 
In contrast to the wide insulating band gap observed in NdFeO$_3$, an increase in the density of states near the Fermi level is observed in NdFeO$_{3-\delta}$ films with higher oxygen vacancy densities resulting from stronger annealing conditions. This observation suggests the presence of low-energy states induced by oxygen vacancies \cite{choi2012anti,tan2013interface,hong2024effect,aiura2002photoemission,meevasana2011creation}. 
Subsequently, the NdFeO$_{3-\delta}$/Nb:SrTiO$_3$ films were transferred to a Chalcogenide-MBE chamber and annealed under Se flux at 950 $^\circ$C for 45 minutes. After the annealing, the valence band peaks of the NdFeO$_3$ films at 4 eV below the Fermi energy are similar to those of the annealed LaFeO$_3$ films under Se flux, suggesting comparable oxygen vacancy densities [see Fig.~\ref{Figure1}(h)]. 
Note that the annealed NdFeO$_{3-\delta}$ films exhibit a higher spectral weight near the Fermi level [see Fig.~\ref{Figure1}(g-h)] , when compared to LaFeO$_{3-\delta}$ \cite{song2021high}. This observation indicates that the concentration of oxygen vacancies in the surface layers of NdFeO$_3$ is no less than that in LaFeO$_3$. 

\begin{figure*}[bht]
\includegraphics{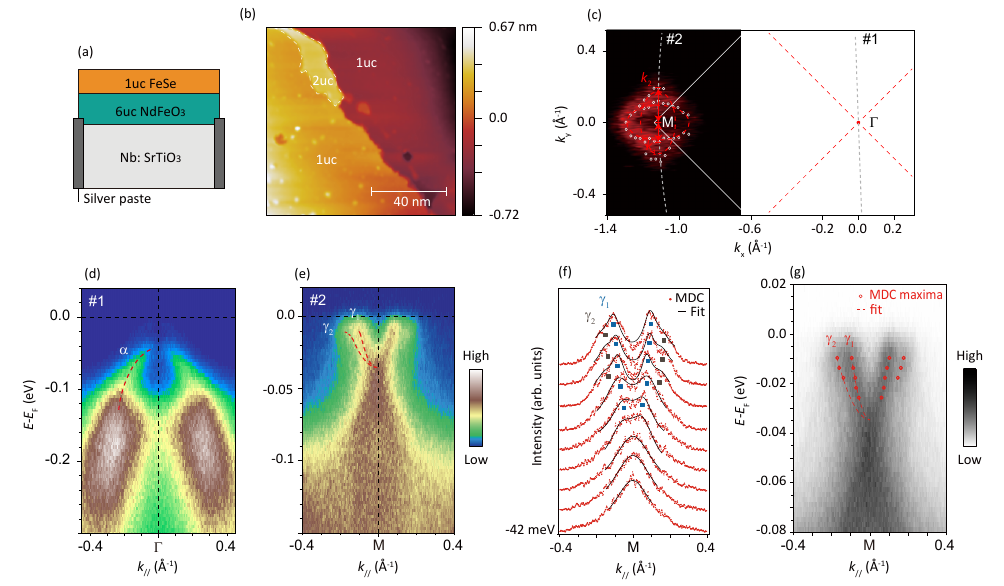}% Here is how to import EPS art
\caption{\label{Figure2}(a) Structure of 1uc-FeSe/6uc-NdFeO$_3$/Nb:SrTiO$_3$. (b) STM morphology of annealed 1uc-FeSe/6uc-NdFeO$_3$/Nb:SrTiO$_3$. (c) Fermi surface of 1uc-FeSe/6uc-NdFeO$_3$/Nb:SrTiO$_3$ heterostructure. Open circles refer to fitted $k_F$ from in-plane mappings (cut by cut) and the grey elliptical circle refer to 2 non-degenerate electron pockets at the M point, suggesing relatively low electron doping level. (d) Band dispersion near $\Gamma$ point along the cut $\#$1 shown in Fig (c). The hole band of FeSe is labeled as $\alpha$.  (e)  Band dispersion near M point along the cut $\#$2 shown in Fig (c), labeling the 2 electron bands as $\gamma_1$ and $\gamma_2$.  (f) MDC near the M point. The FeSe/ NdFeO$_3$ exhibits non-degenerate electron pockets at the Fermi level. (g) Fitted electron bands plotted on the band dispersion along the cut $\#$1 shown in Fig (c). All the ARPES measurements were conducted at 9K.}
\end{figure*}

The topography measurements were performed by a scanning tunneling microscope[STM, (RHK Technology)] that in-situ connected with the ARPES chamber and MBE chambers in the combined ultra-high vacuum (UHV) system. The samples were measured at 17 K under pressure of $\sim$ 1 $\times$ 10 $^{-10}$ mbar.  The annealed NdFeO$_3$/Nb:SrTiO$_3$ films exhibit atomically flat surfaces, as observed by in-situ scanning tunneling microscopy (STM) in Fig.~\ref{Figure1}(i). Subsequently, 1uc FeSe films were grown on the NdFeO$_3$/Nb:SrTiO$_3$ heterostructures at a temperature of 490 $^\circ$C, using a Fe/Se flux rate ratio of 1:10, followed by annealing at 520 $^\circ$C for 3 hours.  The STM morphology image of the annealed 1uc FeSe/NdFeO$_3$/Nb:SrTiO$_3$ [Fig.~\ref{Figure2}(b)] exhibits well-defined terrace edges,
predominantly 1uc in thickness with a few 2~uc areas [Supplementary Fig.~S2]. 

\section{L\MakeLowercase{ow electron doping in 1uc} F\MakeLowercase{e}S\MakeLowercase{e}/6\MakeLowercase{uc} N\MakeLowercase{d}F\MakeLowercase{e}O$_3$/N\MakeLowercase{b}:S\MakeLowercase{r}T\MakeLowercase{i}O$_3$\label{section3}}

In-situ angle-resolved photoemission spectroscopy (ARPES) studies were performed on the grown 1uc-FeSe/6uc-NdFeO$_3$/Nb:SrTiO$_3$ (hereafter referred as FeSe/NFO) with the 21.2eV(He-$\rm\uppercase\expandafter{\romannumeral1}\alpha$) helium discharge lamp from Fermi Instrument and the VG-Scienta DA30 electron analyzer, under the ultra-high vacuum better than 2.5 ${\times}$ 10 $^{-11}$ mbar. All ARPES measurements were conducted at 9K. The energy resolution is 6 meV and the angular resolution is 0.3$^\circ$. Around the $\Gamma$ point, a parabolic band denoted as $\alpha$ band was observed, with its band top locate at 45 meV below the Fermi energy [Fig.~\ref{Figure2}(d)], which is higher than that of FeSe/SrTiO$_3$ of 80 meV below the Fermi energy and FeSe/LaFeO$_3$ of 70 meV below the Fermi energy, suggesting that the electron doping level of the FeSe/NdFeO$_3$ interface is lower than that in FeSe/SrTiO$_3$ interface and FeSe/LaFeO$_3$ interface.

Figure~\ref{Figure2}(e) shows the measured band structure near the M point, where two electron bands, labeled as $\gamma_1$ and $\gamma_2$. The electron bands $\gamma_1$ and $\gamma_2$ can be further identified by extracted momentum distribution curves (MDCs) and fitted bands according to peak positions of MDCS [Fig.~\ref{Figure2}(f-g)]. The sizes of the semi-minor axis ($k_1$) and the semi-major axis ($k_2$) of the elliptical Fermi pocket were determined to be 0.096~$\AA^{-1}$ and 0.172~$\AA^{-1}$, respectively. According to Luttinger's theorem, the electron doping level of the FeSe/NFO interface, derived from the Fermi surface volume, is 0.040 e$^-/\rm{Fe}$, which is lower than that of 1uc FeSe/SrTiO$_3$ interface(0.10$\sim$0.12 e$^-/\rm{Fe}$) and 1uc FeSe/LaFeO$_3$ interface (0.087 e$^-/\rm{Fe}$).  
To confirm this, experiments were repeated on a series of 1uc FeSe/NdFeO$_3$/Nb:SrTiO$_3$ samples, and the electron doping level ranged from 0.038 e$^-/\rm{Fe}$ to 0.046 e$^-/\rm{Fe}$ (refer to supplementary Fig.~S3), indicating a consistently low doping level at the FeSe/NFO interface.  The band structure of the FeSe/NFO is similar to that of the FeSe/SrTiO$_3$ and FeSe/LaFeO$_3$ interfaces, except for a downward shift in the chemical potential \cite{tan2013interface,song2021high}. 

Such a low electron doping level of 1uc FeSe/NdFeO$_3$ falls within the underdoped region of electron-doped FeSe \cite{wen2016anomalous}, suggesting that the charge transfer effect at the FeSe/NdFeO$_3$ interface is weaker compared to that of FeSe/SrTiO$_3$ and FeSe/LaFeO$_3$, despite the presence of an similar quantity of oxygen vacancies as suggested by the valence band structure in Fig.~1(h). 

\section{S\MakeLowercase{uppressed nematicity in 1uc} F\MakeLowercase{e}S\MakeLowercase{e}/6\MakeLowercase{uc} N\MakeLowercase{d}F\MakeLowercase{e}O$_3$/N\MakeLowercase{b}:S\MakeLowercase{r}T\MakeLowercase{i}O$_3$\label{section4}}

\begin{figure}[bht]
\includegraphics{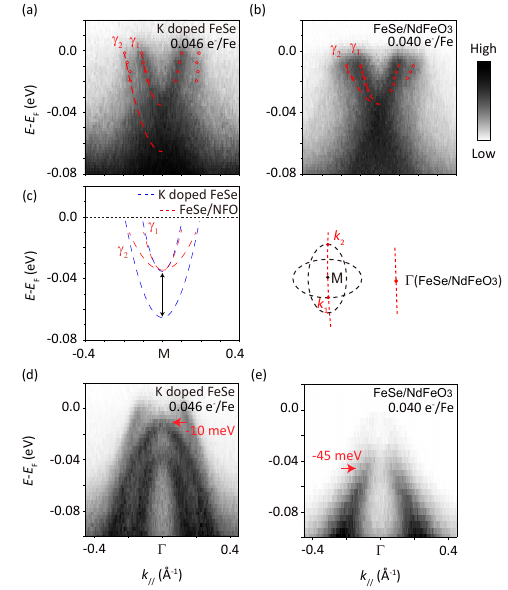}% Here is how to import EPS art
\caption{\label{Figure3}Band dispersion and fitting at M point of different samples: (a) K-doped FeSe with the doping level of 0.046e$^-/\rm{Fe}$; (b) FeSe/NFO with doping level of 0.040 e$^-/\rm{Fe}$. (c) Fitting comparison between K-doped FeSe and FeSe/NFO. (d) Band dispersion of FeSe/NFO at the $\Gamma$ point; (e) Band dispersion of K-doped FeSe films at the $\Gamma$ point. The band tops which do not cross the Fermi level are marked by red arrows.}
\end{figure}

Figure \ref{Figure3} shows the band structure of a surface potassium(K)-doped FeSe sample and the FeSe/NdFeO$_3$ samples in comparable doping levels. Around M, for K-doped FeSe, the band bottom of $\gamma_2$ is at $E_{\rm{F}}$-65 meV and the band bottom of $\gamma_1$ is at $E_{\rm{F}}$-35 meV, showing a 30 meV splitting of the band bottoms at M [Fig. \ref{Figure3}(a,c)]. In contrast, by fitting the peak positions of the MDCs to a parabolic dispersion, we find that the band  $\gamma_1$ and the band $\gamma_2$ of  FeSe/NdFeO$_3$ samples exhibit the same band bottom at $E_{\rm{F}}$-34 meV, without splitting [Fig. \ref{Figure3}(b,d)].
Around $\Gamma$, for K-dosed FeSe films [Fig.3(d)], there is one hole band crossing the Fermi level from the undoped FeSe underlayer, while the topmost doped FeSe shows two hole bands with the band top located at $E_{\rm{F}}$-10 meV for the band most close to the Fermi level \cite{wen2016anomalous}.
However, for FeSe/NdFeO$_3$, the top of hole band at the $\Gamma$ point is located further below the Fermi level ($E_{\rm{F}}$-45 meV), which is the character of 1uc FeSe/oxides~[Fig. 3(e)].
Therefore, there is a stark contrast in the band structure between 1~uc FeSe/NdFeO$_3$ and  K-doped FeSe thick films with comparable electron doping levels (0.040 e$^-/\rm{Fe}$ for FeSe/NdFeO$_3$ and 0.046 e$^-/\rm{Fe}$ for K-doped FeSe thick films~\cite{wen2016anomalous}). 

The splitting of bands $\gamma_1$ and  $\gamma_2$ is commonly observed in FeSe bulk crystals \cite{nakayama2014reconstruction,yi2019nematic,zhang2015observation}, FeSe thick films \cite{wen2016anomalous,miyata2015high,tan2013interface} and underdoped K-dosed FeSe \cite{wen2016anomalous} below the nematic transition temperature, representing different occupation between the  $d_{xz}$ and $d_{yz}$ orbitals due to nematicity \cite{bohmer2022nematicity,rhodes2022fese}. 
The larger energy separation of the bands reflects the stronger strength of the nematic order \cite{tan2013interface}. Here the negligible energy separation in 1~uc FeSe/NdFeO$_3$ suggests an absence of nematicity. It should be noted that such doping corresponding to a nematic regime in thick FeSe films \cite{wen2016anomalous}, and thus the interface plays intriguing roles in varying the band structure and suppressing nematicity beyond simple electron doping.

\begin{figure*}[bht]
\includegraphics{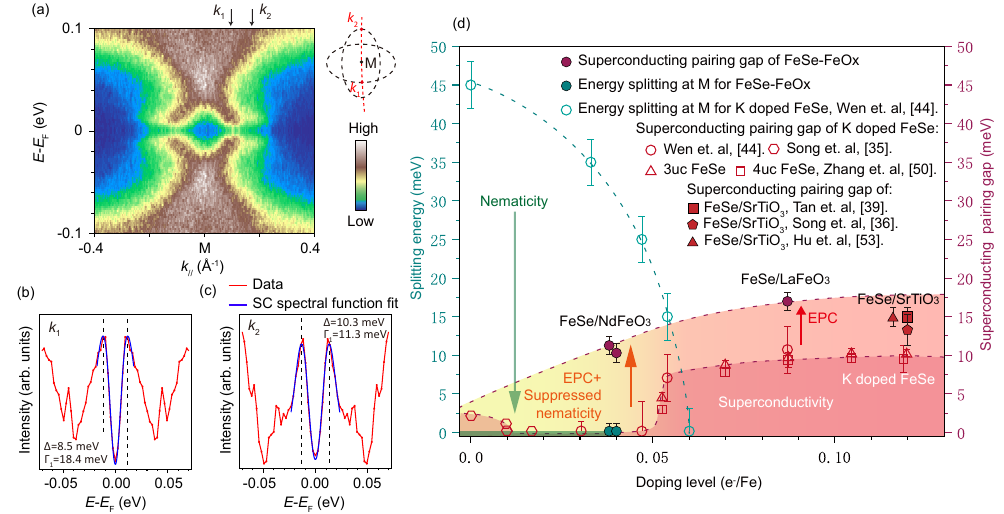}% Here is how to import EPS art
\caption{\label{Figure4}(a) Symmetrized band dispersion of \# 2 shown in Fig. \ref{Figure2}(e), showing the superconducting pairing gap $\Delta$. (b) Energy distribution curve at the momentum of $k_{1}$ shown in Fig (a) with the fitted superconducting pairing gap $\Delta_{k_{1}}$=8.5 meV and the single-particle scattering rate $\Gamma_1$ of 18.4 meV. (c) Energy distribution curve at the momentum of $k_{2}$ shown in Fig (a) with the fitted superconducting pairing gap $\Delta_{k_{2}}$=10.3 meV and the single-particle scattering rate $\Gamma_1$ =11.3 meV. All the ARPES measurements were taken at 9K. (d) Phase diagram of nematic order induced energy splitting and superconducting pairing gap for FeSe-FeO$_x$, FeSe/SrTiO$_3$ and K-doped FeSe against electron doping level. The solid-circle-shaped data refer to FeSe-FeO$_x$. The open-circle-shaped data refer to K-doped FeSe data adapted from ref.~\cite{wen2016anomalous}~(Wen et. al), the open-triangle-shaped data and the open-square-shaped data refer to K-doped FeSe data adapted from ref.~\cite{zhang2016effects}~(Zhang et. al), the open-hexagon-shaped data refer to K-doped FeSe data adapted from ref.~\cite{song2016observation}~(Song et. al), the solid-square-shaped data refer to well-annealed 1uc FeSe/SrTiO$_{3}$ data adapted from ref.~\cite{tan2013interface}~(Tan et. al), the solid-pentagon-shaped data refer to well-annealed 1uc FeSe/SrTiO$_{3}$ data adapted from ref.~\cite{song2019evidence}~(Song et. al) and the solid-triangle-shaped data refer to well-annealed 1uc FeSe/SrTiO$_{3}$ data adapted from ref.~\cite{hu2018identification}~(Hu et. al).  FeSe-FeO$_x$ interfaces exhibit superconducting pairing strength enhancement and nematicity suppression compared to K-doped FeSe.}
\end{figure*}

\section{S\MakeLowercase{uperconducting gap and phase diagram}}

Figure \ref{Figure4}(a) presents the symmetrized photoemission spectrum of 1uc-FeSe/6uc-NdFeO$_3$/Nb:SrTiO$_3$ measured at 9~K at the M point.  The observed back-bending behavior is a hallmark of Bogliubov quasiparticle dispersion, and a superconducting gap opening is observed.
Figure \ref{Figure4}(b) and Fig. \ref{Figure4}(c) show the energy distribution curve~(EDC) and corresponding fits for the superconducting gap at two momentum positions. According to spectral function fitting, the superconducting gap sizes $\Delta$ at $k_{1}$ (the minor axis endpoint of the Fermi pocket at M point) and $k_{2}$ (the major axis endpoint of the Fermi pocket at M point) are 8.5 meV and 10.3 meV respectively.  Furthermore, superconducting spectral function fitting provides insights into the single-particle scattering rate $\Gamma_2$, which reflects the level of impurity scattering and the quality of the films. 
The single-particle scattering rate coefficient $\Gamma_2$ is 11.3 meV at $k_{2}$ (the outer elliptical electron pocket), which is similar to that of the high-quality 1uc FeSe/SrTiO$_3$ heterostructures ($\Gamma_2$ = 12 meV)~\cite{song2019evidence}, indicating that the 1uc FeSe/6uc NdFeO$_3$/Nb:SrTiO$_3$ samples are of good quality without strong impurity scattering.  Note that the superconducting gap is absent in K-doped FeSe films with a comparable doping level \cite{wen2016anomalous,song2016observation}, suggesting that the origin of the enhanced superconductivity in FeSe/NdFeO$_3$ is beyond simple electron doping.

In Fig.4(d), we plot the energy scales of superconducting gap and nematic splitting in 1uc FeSe films interfaced with various FeO$_x$  terminated oxides as a function of electron doping \cite{song2021high}, together with those of the surface doped thick FeSe films \cite{zhang2016effects,wen2016anomalous,song2016observation} and well annealed 1uc FeSe/SrTiO$_3$\cite{tan2013interface, faeth2021interfacial, jia2021magic, liu2021high, peng2014tuning, rebec2017coexistence}. For electron-doped FeSe without an oxide interface, the comparable electron doping level of FeSe/NdFeO$_3$ is close to the boundary of superconductivity and in a partially suppressed nematic order state \cite{wen2016anomalous,song2016observation}.  For 1uc FeSe/NdFeO$_3$
 heterostructure, a superconducting pairing gap $\Delta$ of 10.3 meV is observed with no energy splitting at M point, suggesting that nematicity is suppressed while superconducting pairing strength is enhanced by the  FeSe/NdFeO$_3$ interface. 

\section{D\MakeLowercase{iscussion}}

Previous ARPES studies have revealed that for the 1uc FeSe/TiO$_x$ interface, the doping level of 1uc FeSe films is pined to 0.10$\sim$0.12  e$^-/\rm{Fe}$ (i.e. “magic doping”) \cite{li2022modulation,jia2021magic}. However, the well-annealed 1uc-FeSe/6uc-NdFeO$_3$/Nb:SrTiO$_3$ samples have lower electron doping levels of 0.038$\sim$0.046 e$^-/\rm{Fe}$, deviating from the “magic doping”. Based on the VB shift and density of states (DOS) at E$_F$, it is evident that the density of oxygen vacancies is equal to or exceeds that in FeSe/LaFeO$_3$. Therefore, the NdFeO$_3$ acts as “infinite” charge reservoir, and the doping level of FeSe/NdFeO$_3$ is not constrained by oxygen vacancies but could potentially be limited by a smaller work function mismatch. This calls for future studies into the variations of work functions across different FeSe-FeO$_x$ and FeSe-TiO$_x$ interfaces.

FeSe/NdFeO$_3$ is the first well-annealed FeSe/oxide interface that exhibits low electron doping in 1uc FeSe films, which provides new insights into the role of the interface.
It is shown that nematicity can be enhanced by tensile strain \cite{phan2017effects,tan2013interface,li2017stripes,zhang2016distinctive}. Consequently, it is expected nematicity would be strongest in 1~uc FeSe/STO with the strongest tensile strain if without interfacial electron transfer~\cite{tan2013interface}.
However, our results provide direct evidence that the nematicity is even weaker in 1~uc FeSe/NdFeO$_3$ than the similarly doped thick FeSe with less tensile strain. 
Further theoretical development is required to elucidate the origin of the suppressed nematicity by the interface, while it is likely that the altered band structure of FeSe by the oxide interface plays some role. In iron-based superconductors, nested electron and hole pockets at the Fermi surface are suggested to induce strong spin fluctuation~\cite{hirschfeld2011gap} and drive nematicity \cite{wang2016strong}. When the nesting conditions deteriorate, the nematic order is suppressed \cite{narayan2023potential}. In comparison to K-dosed thick FeSe with similar electron doping, the hole-like band of 1-uc FeSe/NdFeO$_3$ is far from the Fermi surface, and the significantly poorer nesting conditions may contribute to the suppression of nematicity.

The interaction between the electrons in FeSe and interfacial optical photons has been suggested to boost superconductivity in FeSe/STO and FeSe/LaFeO$_3$ \cite{song2021high,song2019evidence,faeth2021interfacial}. There are two optical phonon modes in NdFeO$_3$ corresponding to the relative motion of Fe-O atoms within the oxygen octahedron of 55.8 meV and 79.8 meV, respectively \cite{deliormanli2021erbium}. The 79.8 meV optical phonon mode is similar in energy scale to the previously reported phonon modes in the FeSe/LaFeO$_3$ interface\cite{song2021high}.
Although the data statistics of the EDC at M point are not high enough for replica band fitting, the integrated EDC shows a peak-dip-hump structure M point, and the energy separation between the peak and hump is around 80 meV, supporting the coupling between electrons and the FK phonons of NdFeO$_3$ \cite{song2019evidence,song2021high} (Details can be seen in supplementary Fig.S5). 
Furthermore, given that FeSe/NdFeO$_3$ is closer to the nematic phase but lacks nematic order, it is intriguing to investigate whether nematic fluctuations play an additional role in enhancing superconductivity in the low electron-doped region, in addition to the cooperative effect of interfacial electron-phonon coupling (EPC).

To summarize, we have successfully grown FeSe/NdFeO$_3$ interface, and demonstrate it as a new member of the FeSe/FeO$_x$ interfacial superconductor family. 
Note that there are multiple magnetic ordered states in NdFeO$_3$\cite{treves1965studies,przenioslo2006nuclear,sosnowsk1986reorientation,bartolome1997single,khaled2022spin,khaled2022strain}, and the extending of FeSe/FeO$_x$ interfacial superconductor family provides a new platform to engineer and study the interaction between superconductivity and magnetism. Moreover, the charge transfer of the interface can be controlled at the underdoped region (0.038$\sim$0.046 e$^-/\rm{Fe}$), which is the first underdoped case in the well-annealed 1uc FeSe/oxide family. The observed nematicity suppression provides new information in understanding the interaction between oxide interfaces and FeSe. 

\begin{acknowledgments}
This work is supported in part by the National Science Foundation of China   
 under the grant Nos. 11922403, 12274085, 12074074 and 92365302, the National Key R\&D Program of the MOST of China (2023YFA1406300), the New Cornerstone Science Foundation, the Innovation Program for Quantum Science and Technology (Grant No. 2021ZD0302803), and Shanghai Municipal Science and Technology Major Project (Grant No.2019SHZDZX01). 
\end{acknowledgments}

\appendix

%\bibliography{apssamp}% Produces the bibliography via BibTeX.

\end{document}